\documentclass[aps,prl,twocolumn,showpacs,super.scriptaddress,groupedaddress]{revtex4}  
\usepackage{graphicx}  
\usepackage{dcolumn}   
\usepackage{bm}        
\usepackage{amssymb}   

\usepackage{amsmath,  amssymb,   amsthm,  mathrsfs}
\usepackage{color}
\def\be{\begin{equation}}
\def\ee{\end{equation}}
\def\bea{\begin{eqnarray}}
\def\eea{\end{eqnarray}}
\def\ba{\begin{array}}
	\def\ea{\end{array}}
\newcommand{\bes}{\begin{subequations}}
	\newcommand{\ees}{\end{subequations}}
\newcommand{\sect}[1]{\noindent\textit{#1}.---\!}

\begin{document}

\title{Scale vs Conformal invariance from Entanglement Entropy}

\author{Ali Naseh}
\email{naseh@ipm.ir}

\affiliation{School of Particles and Accelerators, Institute for Research in Fundamental Sciences (IPM)\\
	P.O.Box 19395-5531, Tehran, Iran
}
\begin{abstract}
For a generic conformal field theory (CFT) in four dimensions, the scale anomaly dictates that the universal part of entanglement entropy across a sphere ($\mathcal{C}_{\text{univ}}(S^{2})$) is positive. Based on this fact, we explore the consequences of assuming positive sign for $\mathcal{C}_{\text{univ}}(S^{2})$ in a four dimensional scale invariant theory (SFT). In absence of a dimension two scalar operator $\mathcal{O}_{2}$ in the spectrum of a SFT, we show that this assumption suggests that SFT is a CFT. In presence of $\mathcal{O}_{2}$, we show that this assumption can fix the coefficient of the nonlinear coupling term $\int\hspace{-.5mm} d^{4}x\sqrt{g} R\mathcal{O}_{2}$ to a conformal value.  
\end{abstract}

\maketitle

The asymptotic structure of Poincar\'{e} invariant unitary quantum field theories in deep UV and IR is of great importance in physics. A deep understanding of this issue is achievable via the profound idea of Wilson \cite{Wilson:1973jj}. According to this idea, the fixed points of renormalization group (RG) are dwellings of that asymptotics and therefore the asymptotic theories are scale invariant. Other new dwellings are the renormalization group limit cycles which also describe the scale invariant field theories. Remarkably, with a few known exceptions, unitary SFT's always exhibit full conformal symmetry. A natural question is whether it is possible for a theory to be scale invariant but not conformal invariant? The converse question, i.e., whether a theory can be invariant under conformal transformations but not under scaling, is easy to answer. The commutator between the conserved generators of translations and conformal transformations gives the scaling generator together with the Lorentz ones. This means that Poincar\'{e} plus conformal invariance comprises scale invariance. The converse is still an open question since Poincar\'{e} and scaling generators form a closed algebra.

Recently there were considerable efforts to answer this question and the task has been done in some spacetime dimensions, but the problem is still open for $D=4$. Although some comprehensive arguments are available in 4D, they still suffer from serious loophole. In this paper we study the problem of scale vs conformal invariance in 4D by making use of entanglement entropy. For a generic CFT in 4D, the scale anomaly dictates that the universal part of entanglement entropy across a sphere ($\mathcal{C}_{\text{univ}}(S^{2})$) is positive \cite{Perlmutter:2015vma}. Based on this fact, we explore the consequences of assuming positive sign for $\mathcal{C}_{\text{univ}}(S^{2})$ in a 4D SFT. In absence of a dimension two scalar operator $\mathcal{O}_{2}$ in the spectrum of a SFT, we show that this assumption suggests that the SFT actually is a CFT. In presence of $\mathcal{O}_{2}$, which is actually related to the loophole in previous studies, we show that this assumption fixes the coefficient of the nonlinear coupling term $\int\hspace{-.5mm} d^{4}x\sqrt{g} R\mathcal{O}_{2}$ to a conformal value.  

The paper is organized as follows. The first section is devoted  to a comprehensive review on previous studies on the subject of scale vs conformal invariance by emphasizing on 4D. Since our work is highly based on using scale anomaly in SFTs we will dedicate some parts of the first section to this topic and its crucial rule in the subject of scale vs conformal invariance. Also in this section, the remaining problem in previous studies would be mentioned. In section two we study  scale vs conformal invariance in 4D via entanglement entropy. Finally in the last section, we will discuss the possible appearance of scale anomaly in other measures of entanglement and their rule in the problem of scale vs conformal invariance.

\sect{Previous Attempts on Scale vs Conformal Invariance}
In $D=2$, based on the argument of Zamolodchikov \cite{Zamolodchikov:1986gt}, Polchinski  proved that any unitary SFT exhibits full conformal symmetry \cite{Polchinski:1987dy}. Polchinski assumed that a unitary 2D SFT has a well-defined energy-momentum tensor together with a discrete spectrum and finite energy-momentum two-point function.\hspace{.5mm}Later on Riva and Cardy presented a model with scale but without conformal symmetry \cite{Riva:2005gd}. However their model does not violate Polchinski's argument because it does not have reflection-positivity, the Euclidean version of unitarity, and more precisely it does not have a discrete spectrum. An earlier model by Hull and Townsend \cite{Hull:1985rc} which seems to be in contradiction with Polchinski proof is not also a counterexample because this model violates the assumption of having well-behaved energy-momentum two-point function. More recent proposed counterexamples also violate one of the assumptions of the theorem (unitarity, existence and finiteness of correlators) \cite{Iorio:1996ad,Ho:2008nr}.

For $D \geq 3$, the situation was unclear up to 2011. Actually all the perturbative fixed points, which were introduced in the pre-existing literature, belonged to two general categories. In the first category, the fixed points come from the RG flow in theories which do not have any candidate for virial current and therefore that fixed points were automatically conformal invariant \cite{Belavin:1974gu,Banks:1981nn}. In the second category, which is more interesting, although the studied theories have a nontrivial candidate for virial current, but at the fixed points no virial current appears and therefore they also exhibit full conformal symmetry \cite{Polchinski:1987dy,Dorigoni:2009ra}. Consequently, a general conjecture seemed to be that the Zamolodchikov-Polchinski theorem is even true in $D \geq 3$, even though a proof has not been available. Interestingly in 2011 it was demonstrated that this conjecture is false, at least in $D = 3$ and in $D \geq 5$ \cite{ElShowk:2011gz}. The counterexample is simply the free Maxwell theory. This scale invariant field theory is unitary, it has a well-defined energy-momentum tensor and also has a discrete spectrum, but it is not a CFT. 

Therefore,  we are remained with $D=4$. Firstly, it is shown that at all $4D$ perturbative fixed points the scale symmetry is enhanced to the full conformal symmetry \cite{Luty:2012ww,Baume:2014rla}. Indeed the approach of \cite{Luty:2012ww} is based on the idea of Komargodski-Schwimmer's a-theorem, while  \cite{Baume:2014rla} is based on the concept of local Callan-Symanzik equation. The argument in \cite{Luty:2012ww} holds even for theories with gravitational anomalies. Furthermore, it is argued that perturbative scale-invariant trajectories correspond to rare RG flows, namely limit cycles with non-vanishing beta functions \footnote{According to the work of Jack and Osborn \cite{Jack:1990eb,Osborn:1991gm}, a theory does not need to have zero beta functions in order to be conformal.}, also enjoy the benefit of conformal symmetry \cite{Fortin:2011ks,Fortin:2012ic,Fortin:2012cq,Fortin:2012hc}. 
Furthermore in \cite{Luty:2012ww} it was proposed that scale anomaly can be used to understand the scale vs conformal invariance at non-perturbative level. Anomalies are caused by quantum effects. At the classical level a general SFT has a local conserved scale current $S^{\mu}$ \cite{Wess:1960}
\bea\label{S.Current}
S^{\mu} = x^{\nu}T_{\nu}^{\mu} + V^{\mu},
\eea	
where $T_{\mu \nu}$ denotes the energy-momentum tensor and $V^{\mu}$ is 
the so-called 'virial current'. Conservation of scale current gives
\bea
0 = \partial_{\mu} S^{\mu} = T^{\mu}_{\mu} +\partial_{\mu}V^{\mu},
\eea  
which means that for scale invariant theories $T_{\mu}^{\mu} = -\partial_{\mu}V^{\mu}$. Note
that we have used the fact that the energy-momentum tensor is conserved. Obviously if the virial current in a SFT is conserved, that SFT actually is a CFT. The less obvious case in which a unitary SFT would
be a CFT is when the virial current is a total derivative, i.e, 
\bea\label{nontrivir}
V_{\mu} = \partial_{\mu}L.
\eea
In such a case one can find an improved energy-momentum tensor
\bea
\tilde{T}_{\mu\nu} = T_{\mu\nu}+ 
\frac{1}{3}(\partial_{\mu}\partial_{\nu}-\eta _{\mu\nu}\square) L,
\eea
which is conserved and traceless \cite{Polchinski:1987dy,Dymarsky:2013pqa,Dymarsky:2014zja}. In the following by SFT we mean a theory that its ’virial current’ is neither conserved nor a total derivative. At the quantum level, in general, scale invariance may be broken by anomalies. The anomalies can be represented in terms of the Wess-Zumino action. In order to proceed, a convenient formalism is to introduce background fields $g_{\mu\nu}$ and $C_{\mu}$ as a source for $T_{\mu\nu}$ and $V_{\mu}$ respectively. In this way,
\begin{align}
\nonumber & \hspace{.5cm} e^{i W[g_{\mu\nu},C_{\mu}]} = \int d[\varphi]~e^{i S[\varphi ;g_{\mu\nu},C_{\mu}]}, \\ 
& T^{\mu\nu} = \frac{2}{\sqrt{-g}}\frac{\delta S}{\delta g_{\mu\nu}},~~~~~~~
V^{\mu} = -\frac{1}{\sqrt{-g}}\frac{\delta S}{\delta C_{\mu}},  
\end{align}
where $W$ is the generating functional of connected graphs. Under the generalized Weyl transformation \footnote{We implicitly assumed that the commutator of scale generator with energy-momentum tensor has a  canonical form. For discussion about that assumption see \cite{Polchinski:1987dy,Dymarsky:2013pqa,Osborn:1991gm,Baume:2014rla}.}
\bea\label{GW}
\delta_{\sigma}g_{\mu\nu} = 2\sigma g_{\mu\nu},~~~~~~\delta_{\sigma}C_{\mu} =
\partial_{\mu}\sigma ,
\eea
we have
\begin{align}\label{Del.S.W}
\nonumber & \delta_{\sigma}W = \int d^{4}x \left(
-2\sigma g^{\mu\nu}\frac{\delta}{\delta g^{\mu\nu}}+\partial_{\mu}\sigma
\frac{\delta}{\delta C_{\mu}}\right)W, \\ 
& \hspace{.75cm} = \int d^{4}x\sqrt{-g}\sigma \langle T_{\mu}^{\mu}+\nabla_{\mu}V^{\mu}\rangle.  
\end{align}
If the SFT is non-anomalous, $\delta_{\sigma}W$ vanishes. But in the presence of anomaly in general we have 
\bea
\delta_{\sigma} W = S_{WZ}\big{|}_{\sigma} ,
\eea
which results in
\bea\label{Anomalous.Trace}
\int d^{4}x\sqrt{-g}\sigma \langle T_{\mu}^{\mu}+\nabla_{\mu}V^{\mu}\rangle 
= S_{WZ}\big{|}_{\sigma}.
\eea
Here, $S_{WZ}\big{|}_{\sigma}$ denotes those terms in Wess-Zumino action ($S_{WZ}$) which are linear in $\sigma$. The most general parity even Wess-Zumino action involving the metric and the gauge field $C_{\mu}$ for a 4D SFT is given by \cite{Luty:2012ww,Farnsworth:2013osa}
\begin{align}\label{WZ}
\nonumber & S_{WZ}[g_{\mu\nu},C_{\mu};\sigma]= \int d^{4}x \sqrt{-g}
\bigg\{-a\big[\sigma E_{4}+4(R^{\mu\nu} \\ 
\nonumber & \hspace{1.2cm}-\frac{1}{2}R g^{\mu\nu})\partial_{\mu}\sigma \partial_{\nu}\sigma
-4(\partial\sigma)^{2}\square\sigma +2(\partial\sigma)^{4}\big]+ \\
& \hspace{2.4cm}+c\sigma W^{2}-e\sigma \Sigma^{2} + f\sigma C_{\mu\nu}C^{\mu\nu}\bigg\},
\end{align}
where $E_{4}$ and $W^{2}$ are Euler density and square of Weyl tensor respectively and
\bea
\Sigma = \frac{1}{6}R +\nabla_{\mu}C^{\mu}-C_{\mu}C^{\mu},\hspace{.2cm}
C_{\mu\nu} = \partial_{\mu}C_{\nu}-\partial_{\nu}C_{\mu}.
\eea
 The coefficients $a$ and $c$ are the standard conformal anomaly coefficients of a CFT while the $e$ and $f$ terms appear only in a SFT \footnote{None of the anomalies from (\ref{WZ}) can be removed by adding a local (diffeomorphism-invariant) term to $W$.}. It should be noted that in presence of a dimension two scalar operator $\mathcal{O}_{2}$, the term  $\xi\hspace{-.5mm}\int\hspace{-.5mm} d^{4}x\sqrt{-g} \hspace{.5mm}\Sigma\mathcal{O}_{2}$ can be added to the action which only shifts the anomaly coefficient $e$ \cite{Farnsworth:2013osa} \footnote{A well-known example of this phenomenon is the theory of a free scalar $\phi$, which has a dimension two scalar operator $\mathcal{O}_{2}=\phi^{2}$}.

According to Eq.(\ref{Anomalous.Trace}) and (\ref{WZ}), under the global scale transformations we have 
\begin{align}\label{Anomaly}
\nonumber &  \int d^{4}x\sqrt{-g}\sigma \langle T_{\mu}^{\mu}+\nabla_{\mu}V^{\mu}\rangle \big{|}_{C_{\mu} =0}
=\hspace{-.1cm} \int d^{4}x \sqrt{-g}
\sigma\big(-a E_{4}+ \\ 
& \hspace{4.7cm} +c W^{2}-\tilde{e}R^{2}\big),  
\end{align}
where the normalized $\tilde{e}\hspace{-1mm}\equiv\hspace{-1mm}\frac{e}{36}$ is introduced. The $e$-anomaly plays a crucial role in the problem of scale vs conformal invariance at non-perturbative level which can be understood as follows. From (\ref{Anomaly}), the two-point function of the trace of energy-momentum tensor in a 4D flat anomalous scale invariant theory is given by \cite{Farnsworth:2013osa}\footnote{
	We have implicitly assumed that a SFT has local excitations. In topological QFTs
	the $\tilde{e}$ does not appear \cite{Witten:1988ze}.}
\bea\label{TT}
\langle T(q)T(-q)\rangle = -\tilde{e} q^{4}\log\frac{q^{2}}{\mu^{2}} +B(\mu) q^{4},
\eea
where $\mu$ is an arbitrary renormalization scale and $B(\mu)$ is a scheme dependent constant. It is shown that unitarity imposes 
$\tilde{e} \geq 0$ \cite{Farnsworth:2013osa,Bzowski:2014qja}. Note that the Fourier transformation of the $q^{4}$ term 
in (\ref{TT}) is a delta function, so if
$\tilde{e} = 0$ we have
\bea
\langle T(x)T(0) \rangle =0,~~~~~~~ x\neq 0.
\eea
This means that in a unitary theory, $T$ must be equal to zero as an operator identity and the scale invariant theory becomes fully conformal. It should be noted that to have a CFT in presence of $\mathcal{O}_{2}$, $\tilde{e}$ is not necessarily zero and should satisfy another condition \cite{Bzowski:2014qja}. When this condition holds one may improve $T$ such that the new $T$ vanishes. Based on these observations, it was argued that the structure of a special anomalous 3-point function in any SFT is not compatible with operator product expansions (OPEs) and this implies that the $e$ term must vanish and thus all unitary SFTs are CFTs \cite{Farnsworth:2013osa}. Later on the authors of \cite{Bzowski:2014qja} pointed out a subtlety in the relation between OPEs and the large momentum limit which invalidates this argument. While the OPE controls the leading non-local contribution in the large momentum limit, there are semi-local contributions which dominate over the OPE contribution in the relevant case and therefore the statement in \cite{Farnsworth:2013osa} is false. After that, based on the proof of the a-theorem and using the concept of dilaton scattering amplitudes, it is argued that unitary SFTs must be either CFTs, or the trace of the energy-momentum tensor behaves like a generalized free field \cite{Dymarsky:2013pqa}. Moreover, it is shown that if no scalar operator of dimension precisely 2 appears in the spectrum of a SFT{\color{blue},} which it's energy-momentum tensor is generalized free field, that theory would be conformal \cite{Dymarsky:2014zja}. In the presence of a scalar operator with dimension precisely 2, which can mix with $T$, one can show that there is at least one improvement such that improved $T$ is not a generalized free field \cite{Dymarsky:2014zja}\footnote{We thank Z.Komargodski for a discussion on this issue.}. Thus the only loophole which is remained in the proof of \cite{Dymarsky:2014zja} is the case where the energy-momentum tensor is generalized free field and the scalar operator with dimension precisely 2 exists in the spectrum
\footnote{Recently, it is argued that in any number of spacetime dimensions a SFT embedded inside a unitary CFT must be a free field theory \cite{Dymarsky:2015jia}. Of course, this doesn't mean that the nontrivial SFTs do not exist.}.

In the next section we explore some consequences of assuming positive sign for $\mathcal{C}_{\text{univ}}(S^{2})$ in the subject of scale vs conformal invariance specially in the case where a dimension 2 scalar operator exists in the spectrum of a SFT.

\sect{Entanglement Entropy and Scale vs Conformal}\hspace{.7cm}The properties of non-local quantities are important as the correlation functions of local operators in a given quantum field theory. In particular they are important for understanding of quantum phase structures. One of the important non-local physical quantities is the Wilson loop operator in gauge theories, which is a very useful order parameter for understanding of the confinement \cite{Wilson:1974sk}. Quantum Entanglement (QE) is also a momentous non-local quantity in more generic QFTs. QE has an increasingly dominant impress in understanding of the quantum complex systems in a diverse set of areas including condensed matter physics \cite{Levin:2006zz,Kitaev:2005dm,Li:2008kda,Flammia:2009,Hastings:2010zka}, quantum information theory \cite{Bennett:2000,Nielsen:2000,Vedral:2006}, and quantum gravity \cite{Jacobson:1995ab,Ryu:2006bv,Hubeny:2007xt,VanRaamsdonk:2010pw,Bianchi:2012ev,Maldacena:2013xja,Dong:2013qoa,Dong:2016eik}. One of the measures of QE is entanglement entropy (EE). Considering a pure state of a relativistic SFT defined on a 3+1 dimensional manifold $\mathcal{M}$, EE is defined by tracing out those modes which reside outside an entangling region $\Upsilon$. This entangling region is a submanifold of $\mathcal{M}$ at a fixed time. The result of the trace-out action is a mixed state $\rho _{\Upsilon}$. In order to calculate EE one should first obtain the  $Tr_{\Upsilon}(\rho_{\Upsilon}\hspace{.15mm}^{n})$ and find the R\'{e}nyi entropy
\bea
S_{n}(\rho_{\Upsilon})=\frac{1}{1-n} \log{Tr_{\Upsilon}(\rho_{\Upsilon}\hspace{.15mm}^{n})},
\eea
where $n$ is a positive integer. Upon analytically continuing $n$ to positive real values, one can take the limit $n \rightarrow 1$ to obtain the entanglement, or von Neumann entropy as
\bea\label{EE}
S_{EE} = \lim_{n \rightarrow 1} S_{n} = -\partial_{n}
\log{Tr_{\Upsilon}(\rho_{\Upsilon}\hspace{.15mm}^{n})} \big{|}_{n=1}.
\eea
Furthermore, the $Tr_{\Upsilon}(\rho_{\Upsilon}\hspace{.15mm}^{n})$ can be computed from the partition function $Z_{n}$ 
on a n-sheeted 3+1 dimensional manifold $\mathcal{M}_{n}$ as
\bea\label{TR.Z}
\log{Tr_{\Upsilon}(\rho_{\Upsilon}\hspace{.15mm}^{n})} =\log{Z_{n}}-n\log{Z_{1}}.
\eea
Thus Eq.(\ref{EE}) becomes
\bea\label{EE1}
S_{EE} =  -\partial_{n}\left(\log{Z_{n}}-n\log{Z_{1}}\right) \big{|}_{n=1}.
\eea
For the closed connected surface $\Upsilon$, we can define a length scale $s$. Therefore by using Eq.(\ref{GW}) 
together with (\ref{TR.Z}) and (\ref{Del.S.W}) we have
\bea
&&s\frac{d}{ds}\log{Tr_{\Upsilon}(\rho_{\Upsilon}\hspace{.15mm}^{n})}
= \int_{\mathcal{M}_{n}}\hspace{-.4cm}d^{4}x\sqrt{-g}\langle T_{\mu}^{\mu}+\nabla_{\mu}V^{\mu}\rangle\big{|}_{C_{\mu} =0}\cr\nonumber\\
&&\hspace{1.8cm}-n\int_{\mathcal{M}_{1}}\hspace{-.4cm}d^{4}x\sqrt{-g}\langle T_{\mu}^{\mu}+\nabla_{\mu}V^{\mu}\rangle\big{|}_{C_{\mu} =0}.
\eea
The above result together with (\ref{EE1}) and (\ref{Anomaly}) gives
\bea\label{sdsS}
&&s\frac{d}{ds}S_{EE} =-\partial_{n} \int_{\mathcal{M}_{n}}\hspace{-.4cm} d^{4}x\sqrt{-g}\left(-aE_{4}+cW^{2}-\tilde{e}R^{2}\right)\big{|}_{n=1}\cr\nonumber\\
&&\hspace{1.75cm}+\int_{\mathcal{M}_{1}}\hspace{-.4cm}d^{4}x \sqrt{-g} \left(-aE_{4}+cW^{2}-\tilde{e}R^{2}\right).
\eea
The n-sheeted 3+1 dimensional manifold $\mathcal{M}_{n}$, in general contains conical singularities. The
procedure of calculating the integral of metric curvatures on manifolds with conical singularities 
has been developed in \cite{Fursaev:1995ef,Fursaev:2013fta}. According to that procedure, we have
\bea\label{Sq}
&&\int_{\mathcal{M}_{n}}\hspace{-.4cm}d^{4}x\sqrt{-g} E_{4} = n\int_{\mathcal{M}_{1}}\hspace{-.4cm}d^{4}x\sqrt{-g} E_{4}+8\pi(1-n)\cr\nonumber\\
&&\hspace{2cm}\int_{\partial\Upsilon}\hspace{-.3cm}d^{2}\chi\sqrt{-\gamma} R[\gamma]+\mathcal{O}(1-n)^{2}\cr\nonumber\\
&&\int_{\mathcal{M}_{n}}\hspace{-.4cm}d^{4}x\sqrt{-g} W^{2} = n\int_{\mathcal{M}_{1}}\hspace{-.4cm}d^{4}x\sqrt{-g} W^2+8\pi(1-n)\cr\nonumber\\
&&\hspace{2cm}\int_{\partial\Upsilon}\hspace{-.3cm}d^{2}\chi\sqrt{-\gamma}
K[g;t,s;\mathcal{K}_{ij}^{\alpha}]+\mathcal{O}(1-n)^{2},\cr\nonumber\\
&&\int_{\mathcal{M}_{n}}\hspace{-.4cm}d^{4}x\sqrt{-g} R^{2} = n\int_{\mathcal{M}_{1}}\hspace{-.4cm}d^{4}x\sqrt{-g} R^{2} +8\pi(1-n)\cr\nonumber\\
&&\hspace{2cm}\int_{\partial\Upsilon}\hspace{-.3cm}d^{2}\chi\sqrt{-\gamma} R[g]+\mathcal{O}(1-n)^{2},
\eea
where
\bea\label{K}
K[g;t,s;\mathcal{K}_{ij}^{\alpha}] = 2W_{\mu\nu\alpha\beta}t^{\mu}s^{\nu}t^{\alpha}s^{\beta}-[\mathcal{K}_{ij}^{\alpha}\mathcal{K}^{\alpha ij}
-\frac{1}{2}(\mathcal{K}_{i}^{\alpha i})^{2}],\nonumber\\
\eea
and $g$ is the full 4D metric. Furthermore, $\gamma_{ij}$ and 
$\mathcal{K}_{ij}^{\alpha}$ are the intrinsic metric 
and the extrinsic curvature of $\partial\Upsilon$, $\alpha=
\{t,s\}$ indexing the two normal directions (one timelike
$t^{\mu}$ and one spacelike $s^{\mu}$) and the first term on the
right hand side of (\ref{K}) is nothing but the pullback of the 
Weyl tensor onto $\partial\Upsilon$. Using the relations (\ref{Sq})
in (\ref{sdsS}) one arrives at \footnote{In \cite{Fursaev:2013fta}, the same expression for logarithmic divergent part of EE is deduced from the surface heat kernel coefficient for non-conformal massless field theories. We also thank A.Dymarsky for informing us about previous studies \cite{Banerjee:2014daa,Banerjee:2014hqa} on relation between entanglement entropy and scale vs conformal invariance. We would like to notice that in \cite{Banerjee:2014daa,Banerjee:2014hqa} the authors have just suggested a possible relevance between entanglement entropy and scale vs conformal invariance with no direct evidence. Of course they have not computed the explicit form of the contribution of $e$-anomaly in $\mathcal{C}_{\text{univ}}$.}
\begin{align}\label{sdsEE}
\nonumber & s\frac{d}{ds}S_{EE} = -8\pi \int_{\partial\Upsilon}\hspace{-.3cm}d^{2}\chi\sqrt{-\gamma}\hspace{.5mm}\big(aR[\gamma]-cK[g;t,s;\mathcal{K}_{ij}^{\alpha}]+ \\ 
&\hspace{4.7cm}+\tilde{e}R[g]\big). 
\end{align}
The right hand side of (\ref{sdsEE}) in the absence of the $e$ term is indeed the Graham-Witten anomaly \cite{Graham:1999pm} for a 2 dimensional submanifold $\partial\Upsilon$ on the D dimensional CFT \cite{Schwimmer:2008yh}. The holographic realization of these anomalies comes from studying the Einstein spaces in the bulk which are asymptotically locally AdS manifolds (AlAdS). The former statement means that in the presence of the $e$ term, the right hand side of (\ref{sdsEE}) could be considered as (generalized) Graham-Witten anomalies for a 2 dimensional submanifold on the D dimensional SFT. To check this proposition one could redo the machinery of Graham and Witten for non-AlAdS manifolds such as geometries in the foliation preserving diffeomorphic theory of gravity \cite{Nakayama:2012sn}. 

The point which should be stressed here is that $S_{EE}$ is a UV divergent quantity in a continuum QFT. It has a universal part ($\mathcal{C}_{\text{univ}}$), which is defined as its cutoff-independent terms, which contains non-trivial physical information, including central charges and RG monotones \cite{Solodukhin:2008dh,Myers:2010tj,Casini:2011kv,Jafferis:2011zi}. Furthermore, $s\frac{d}{ds}S_{EE}$ is equal to the minus of $\mathcal{C}_{\text{univ}}$ \cite{Ryu:2006ef}. In many respects, these universal terms are the natural counterparts of quantum-mechanical entropies, which suggest that, in QFT, the \hspace{.5mm}$\mathcal{C}_{\text{univ}}$ is also positive-definite. Indeed, for spherical entangling surfaces ($\partial\Upsilon = S^{2}$) in the vacuum state of CFTs in flat (conformally flat) spacetime this appears to be true \cite{Perlmutter:2015vma,Myers:2010tj, Casini:2011kv,Jafferis:2011zi}\footnote{We thank M.Alishahiha for a discussion on this issue.}. Note that one can always pick complex enough entangling surfaces to violate this positivity \cite{Perlmutter:2015vma}. In this paper, we would like specially to study the effect of $e$-anomaly on the sign of $\mathcal{C}_{\text{univ}}(S^{2})$. For simplicity, we take a conformally flat metric, $g_{\mu\nu}=e^{-2\tau}\eta_{\mu\nu}$ as a background metric. Because the $K[g;t,s;\mathcal{K}_{ij}^{\alpha}]$
is Weyl invariant, it does not contribute to $\mathcal{C}_{\text{univ}}(S^{2})$. Moreover by noting that
\bea
\int_{S^{2}}\hspace{-.2cm}d^{2}\chi\sqrt{-\gamma} 
R[g]\big{|}_{g=e^{-2\tau}\eta}\hspace{-.25cm}\hspace{-.5mm}= 
6\hspace{-1mm}\int_{S^{2}}\hspace{-.3cm}d^{2}\chi\sqrt{-\gamma_{\eta}}\big[
\square\tau-(\partial\tau)^{2}\big],
\eea
from (\ref{sdsEE}) we have  
\bea\label{Cuniv}
\mathcal{C}_{\text{univ}}(S^{2}) = 16\pi \big(a+3
\tilde{e}\int_{S^{2}}\hspace{-.3cm}d^{2}\chi\sqrt{-\gamma_{\eta}}\big[
\square\tau-(\partial\tau)^{2}\big]\big).
\eea
Remember that in a unitary SFT, $\tilde{e}\geq 0$. By assuming $\tilde{e} >0$, one can check that for any positive value of $a$, there exists a function $\tau$ for which the $\mathcal{C}_{\text{univ}}(S^{2})$ becomes negative. On the other hand if $\mathcal{C}_{\text{univ}}(S^{2})$ is a measure for number of degrees of freedom, it can not be negative. Therefore the only possible case to have a positive value for $\mathcal{C}_{\text{univ}}(S^{2})$ is $\tilde{e} =0$. Thus, in absence of a dimension two scalar operator $\mathcal{O}_{2}$ in the spectrum of a SFT, we have shown that positivity of $\mathcal{C}_{\text{univ}}(S^{2})$ suggests that a SFT is a CFT. 
 
Furthermore, as we mentioned in the previous section, the only loophole in the proof of \cite{Dymarsky:2014zja} is related to the case where the trace of energy-momentum tensor is generalized free field and a scalar operator with dimension precisely 2 exists in the spectrum. Also we noted that, in presence of $\mathcal{O}_{2}$, one can add the term $\xi\hspace{-.5mm}\int\hspace{-.5mm} d^{4}x\sqrt{g} R\mathcal{O}_{2}$ to the action in order to change the trace of energy-momentum tensor. This means that the universal part of EE can be changed by adding this nonlinear coupling term \footnote{We thank A.Dymarsky and Y.Nakayama for lots of discussions on this issue.}. To be more precise, this non-linear term just shifts the e anomaly coefficient \cite{Farnsworth:2013osa} in eq.(\ref{Cuniv})
\begin{align}
\nonumber & \mathcal{C}_{\text{univ}}(S^{2}) = 16\pi \bigg(a+3
(\tilde{e}-\alpha\xi)\int_{S^{2}}\hspace{-1mm}d^{2}\chi\sqrt{-\gamma_{\eta}}\hspace{1mm}\times \\ 
&\hspace{4.7cm}\times\big[
\square\tau-(\partial\tau)^{2}\big]\bigg),
\end{align}
where $\alpha$ is a positive number. For example for a free scalar theory, the universal part of EE is calculated by using heat Kernel method \cite{Fursaev:2013fta} which leads to $\tilde{e}=\frac{1}{72}$ and $\alpha=\frac{1}{12}$. Interestingly, positivity of $\mathcal{C}_{\text{univ}}(S^{2})$ fixes the coefficient of non-linear coupling term to $\xi=\frac{\tilde{e}}{\alpha}$ where for free scalar theory becomes $\xi=\frac{1}{6}$. This value for $\xi$ is exactly the one to have a conformal scalar theory. This means that in free scalar theory, the positivity of $\mathcal{C}_{\text{univ}}(S^{2})$ suggests that the theory can be improved to a CFT. 

\sect{Discussion} In the previous section, we have shown that the existence of $e$-anomaly can affect the sign of $\mathcal{C}_{\text{univ}}(S^{2})$, which plays a crucial role in the subject of scale vs conformal invariance in $D=4$. For a generic CFT in four dimensions, the scale anomaly dictates that the $\mathcal{C}_{\text{univ}}(S^{2})$ is positive. Based on this fact, we have explored the consequences of assuming positive sign for $\mathcal{C}_{\text{univ}}(S^{2})$ in a four dimensional SFT. In the absence of a dimension two scalar operator $\mathcal{O}_{2}$ in the spectrum of a SFT, we have shown that this assumption suggests that SFT is a CFT. In the presence of $\mathcal{O}_{2}$ in a SFT, we have shown that this assumption fixes the coefficient of the nonlinear coupling term $\int\hspace{-.5mm} d^{4}x\sqrt{g} R\mathcal{O}_{2}$ to a conformal value. 

The $e$-anomaly may have an effect on strong subadditivity (SSA) inequalities. SSA inequalities state that, given a tripartite quantum system $A,B,C$ and a joint density matrix $\rho(ABC)$, the EEs of the subsystems obey the following inequalities:
\begin{align}
\nonumber & S_{EE}(AB) +S_{EE}(BC) -S_{EE}(ABC) -S_{EE}(B) \geq 0, \\ 
& S_{EE}(AB) +S_{EE}(BC) -S_{EE}(A)-S_{EE}(C) \geq 0.  
\end{align}
SSA is a general theorem that depends only on basic facts about Hilbert spaces and the definition of the von Neumann entropy \cite{Hubeny:2007xt}. It is obeyed as long as the bulk spacetime satisfies the null energy condition (NEC) \cite{Allais:2011ys,Callan:2012ip}. In general, it is believed that the NEC is related to unitarity \cite{Myers:2010tj}. Therefore if in the presence of $e$-anomaly SSA inequalities are violated, the theory is non-unitary and therefore any unitary 4D SFT is a CFT. 

The $e$-anomaly can also affect other measures of QE. For a mixed state the EE is no longer a good measure of entanglement since it mixes quantum and classical correlations. An interesting computable measurement of entanglement for the mixed states is the logarithmic negativity (LN) \cite{Vidal:2002zz,Audenaert:2003,Plenio:2005}, which gives an upper bound on distillable entanglement in quantum mechanics, and is thus strictly greater than the EE. It is argued that 
the universal part of LN is also related to the scale anomalies and for CFTs it is positive definite across spherical entangling surfaces \cite{Perlmutter:2015vma,Calabrese:2012nk,Calabrese:2013mi,Rangamani:2014ywa}. Therefore a natural question would be what happens to the sign of universal part of LN in the presence of $e$-anomaly? To answer this question one should calculate R\'{e}nyi entropies in SFTs. This might be done using the method of \cite{Fursaev:2012mp}. 

The $e$-anomaly may also appear in non-local measures of Quantum Phase Transitions (QPT). One of these non-local measures is EE. In the vicinity of QPTs, EE obeys a scaling behavior \cite{Osterloh:2002,Osborne:2002,Vidal:2003} and its universal properties has been investigated in a family of models \cite{Osterloh:2002,Osborne:2002}. Many other studies of different measures of QPTs have been presented recently. For example QPTs are characterized in terms of the overlap (fidelity) function between two ground states obtained for two close values of external parameters \cite{Zanardi:2007,Zhou:2007,Zhou:2008}.
At the critical point, fidelity shows a peak. This overlap suggests that fidelity may capture some information about finite size scaling and universality classes. Interestingly the holographic counterpart of the fidelity is proposed very recently in \cite{MIyaji:2015mia,Alishahiha:2015rta}. For sure studying the effect of $e$-anomaly on critical exponents and comparing them with simulations may help us to have a better understanding of scale vs conformal invariance. 

\sect{Acknowledgment} Special thank to M. Alishahiha and Z. Komargodski for discussions and encouragements. I have also greatly profited from discussions with A. F. Astaneh, A. Castro, A. Dymarsky , R. Fareghbal, D. Grumiller, A. Mollabashi, M. R. Mohammdai Mozaffar, Y. Nakayama, F. Omidi, S. Rahimi-Keshari, M. Rangamani, S. Rouhani, A.A. Saberi, 
M. M. Sheikh-Jabbari, S. F. Taghavi, M. R. Tanhayi and Yang. Zhou.

\end{document}